\documentclass[%
reprint,
amsmath,amssymb,
aps,
prd,
nofootinbib, floatfix
]{revtex4-2}

\usepackage{graphicx}    
\usepackage{dcolumn}     
\usepackage{bm}          
\usepackage{hyperref}    
\usepackage{xcolor}      
\usepackage{amsfonts}
\usepackage{mathtools}
\usepackage{subcaption}
\usepackage{booktabs}
\usepackage{orcidlink}
\allowdisplaybreaks

\hypersetup{
colorlinks=true,
citecolor=blue,
linkcolor=blue,
urlcolor=blue
}

\graphicspath{{plots/}}

\newcommand{\mpl}{M_{\mathrm{Pl}}}

\begin{document}

\title{Impact of the Infrared Cutoff on Structure Formation in Tsallis Holographic Dark Energy}

\author{Biswajit Das\orcidlink{0000-0002-3350-8594}}
\email{Email: bishoophy@gmail.com}
\affiliation{Physics and Applied Mathematics Unit, \\ Indian Statistical Institute, \\ 203, B. T. Road, Kolkata -- 700 108, India}

\date{\today}

\begin{abstract}

We investigate the viability of Tsallis holographic dark energy (THDE) models, focusing on the role of the infrared (IR) cutoff in the growth of cosmic structures. Considering two commonly used choices of the cutoff, the particle horizon and the future event horizon, we analyze the evolution of linear matter perturbations and compute the growth factor, growth rate, and the observable $f\sigma_8(z)$. These predictions are compared with observational data from redshift-space distortion measurements. We find that the growth history is highly sensitive to the choice of IR cutoff. Models based on the future event horizon are consistent with observational data and can provide a fit comparable to, or slightly better than, the $\Lambda$CDM model for suitable values of the Tsallis parameter $\delta$. In contrast, models constructed using the particle horizon generally fail to reproduce the observed growth of structure. These results demonstrate that the viability of THDE models depends crucially on the choice of IR cutoff and highlight the importance of structure formation as a stringent test of generalized holographic dark energy scenarios.

\end{abstract}

\maketitle

% ==================================================
\section{Introduction}
\label{sec:intro}

The $\Lambda$CDM model has been remarkably successful in explaining a wide range of cosmological observations including the cosmic microwave background (CMB), large-scale structure (LSS),  and baryon acoustic oscillations (BAO). Within this framework, the late-time accelerated expansion of the Universe is attributed to the cosmological constant $\Lambda$, while the formation of large-scale structure is driven by the cold dark matter (CDM). 

Despite its empirical success, the $\Lambda$CDM paradigm faces several longstanding theoretical challenges. The  most prominent ones are the coincidence problem and the cosmological constant problem \cite{lambdaweinberg}. Both of these issues are closely related to our limited understanding of the  microscopic origin of $\Lambda$. Recent observations have also revealed tensions in key cosmological parameters, such as the Hubble constant and the amplitude of matter fluctuations ($\sigma_8$), which may hint at physics beyond the standard model \cite{cosmoverse, sunny}. This motivates us to explore alternative models of dark energy. 

Holographic dark energy (HDE) models provide an alternative in this context. Based on the holographic principle, these models connect the dark energy density to a characteristic infrared (IR) cutoff scale. This leads to a dynamical equation of state of dark energy. While holographic dark energy models can alleviate some conceptual issues of a cosmological constant, their phenomenology is highly sensitive to the choice of the characteristic length scale. 

Holographic dark energy was first proposed in \cite{li}. Since then, a broad class of HDE models has been explored in the literature, typically constructed using an entropy-area relation together with an infrared (IR) cutoff  (see \cite{wang2} for a review). The phenomenology of these models has been explored in, for example,  \cite{horvat, huang, pavon,nojiri1, nojiri2, nojiri3, nojiri4,rocco,tavayef,mohammadi,chimento, chimento2, chimento3, chimento4, chimento5, saridakis, saridakis2, saridakis3, dixit, setare,chakraborty, adhikary, niki, motaghi, luciano, nandhida, srivastava, oliveros, john, manosh, manoshgo, manoshrenyi, guin, pedro, sudipta} and the models have been tested using observational data in, for example, \cite{xin, zhang2, xu, xu2, dheepika, fotios, asghari, leon, purba, ilim, tnl}. While most of the studies focus on the evolution of background quantities, the growth of structures in these models has been previously studied in \cite{bita, tepilakov, das5}. Although the growth of structures in HDE models has been investigated in several contexts, the specific role of the infrared (IR) cutoff in shaping the growth history within Tsallis holographic dark energy (THDE) remains unclear. Since different choices of the IR cutoff can lead to similar background evolution but potentially distinct predictions for structure formation, it is especially important to examine whether these models can consistently reproduce the observed growth of large-scale structure.

We investigate the impact of the IR cutoff on the growth of cosmic structures in THDE models. We consider two commonly used choices of the cutoff, the particle horizon and the future event horizon, and analyze their implications for the evolution of linear matter perturbations. By computing the growth factor, growth rate, and the observable $f\sigma_8(z)$, we compare the theoretical predictions with observational data from redshift-space distortion measurements. This allows a direct assessment of the viability of these models and determine which choice of IR cutoff is favored by structure formation data. To our knowledge, a systematic analysis of structure growth in THDE models with different IR cutoffs, directly confronted with $f\sigma_8$ observations, is limited. This motivates a systematic comparison of different IR cutoffs in THDE models using structure formation observables, particularly $f\sigma_8(z)$.

Unlike many previous studies in which the Tsallis entropy exponent is treated as a free parameter, we fix it to representative values. This isolates the effect of the IR cutoff on the growth of structure more clearly. The Tsallis entropy exponent is associated with the non-extensive nature of gravitational systems and may reflect the non-ergodic behavior of long-range interactions. Fixing this parameter provides a controlled setting to examine the phenomenological implications of the IR cutoff. 

The paper is organized as follows. In Sec.~\ref{sec:hde} we outline the general framework of holographic dark energy and review the Tsallis entropy formalism. We then describe the background cosmology and construct the corresponding THDE. In Sec.~\ref{sec:growth} we present the formalism of the growth of structure in the Universe. The results of our analysis are shown in Sec.~\ref{sec:results}, and we conclude in Sec.~\ref{sec:discussion} with a summary and discussion of the implications of our findings.     

% ==================================================
\section{Holographic Dark Energy}
\label{sec:hde}

\subsection{General Framework for Holographic Dark Energy}

Black hole thermodynamics suggests a radical idea: the number of fundamental degrees of freedom in a gravitational system does not scale with its volume, as one might expect, but rather with the area of its boundary. This is captured by the Bekenstein bound, which states that for a system of total energy $E$ and size $L$, the entropy satisfies $S \leq E L$ \cite{jacob}. In the case of a black hole, this bound is saturated, and the entropy scales as $S_{BH} \sim L^2 \mpl^2$, reflecting an area-law behavior.

This insight led to the formulation of the holographic principle by \cite{gerard} and \cite{susskind}. The central idea is that a consistent theory of gravity should encode all the information within a region in terms of degrees of freedom living on its boundary, implying a general bound $S \leq L^2 \mpl^2$.

However, this picture appears to be in tension with the expectations from ordinary quantum field theory. For a system with a UV cutoff $\Lambda$, the number of degrees of freedom scales extensively, leading to an entropy $S \sim L^3 \Lambda^3$. For sufficiently large regions, this would exceed the holographic bound, suggesting that such a description cannot remain valid in the presence of gravity.

A natural way to reconcile this is to require that the total energy in a region of size $L$ should not be large enough to form a black hole. Identifying the energy as $E = \rho_{\Lambda} L^3$ and imposing the condition $E \leq L \mpl^2$, one obtains an upper bound on the vacuum energy density,
\begin{equation}
    \rho_{\Lambda} \le 3 c^2 \mpl^2 L^{-2},
    \label{eq:rho_l}
\end{equation}
where $c$ is a dimensionless parameter of order unity. If this bound is saturated, the resulting energy density provides the basis for holographic dark energy models.

The above argument can be recast in a more general and illuminating form by directly relating the vacuum energy density to the entropy of the system. Starting from the Bekenstein bound $S \leq E L$ and identifying the total energy as $E = \rho_{\Lambda} L^3$, we obtain
\begin{equation}
    S \leq \rho_{\Lambda} L^4.
    \label{eq:ent_rho}
\end{equation}
Assuming that this bound is saturated, one finds a simple relation between the energy density and the entropy,
\begin{equation}
    \rho_{\Lambda} \sim \frac{S}{L^4}.
    \label{eq:rho_s}
\end{equation}

This expression highlights a key aspect of holographic dark energy: the vacuum energy density is not determined independently, but is instead governed by the underlying entropy associated with the system. Consequently, different assumptions about the entropy lead to different forms of dark energy.

For instance, using the standard Bekenstein--Hawking entropy $S \sim L^2 \mpl^2$ immediately yields
\begin{equation}
    \rho_{\Lambda} \sim \mpl^2 L^{-2},
    \label{eq:hde_s}
\end{equation}
which reproduces the conventional holographic dark energy density. However, if the entropy deviates from the area law—as may occur in systems with long-range interactions or non-standard thermodynamic behavior—the resulting energy density will be modified accordingly.

This observation provides the basis for generalized holographic dark energy models, where alternative entropy formalisms are employed. In particular, adopting Tsallis entropy, which modifies the scaling of entropy with system size, leads to a new class of models with distinct cosmological implications.

\subsection{Tsallis Entropy}

Tsallis entropy is a generalization of the standard Boltzmann--Gibbs (BG) entropy, defined as $S_{BG} = - \sum_i p_i \log p_i$, where $p_i$ denotes the probability of the $i$--th microstate. While the BG entropy is appropriate for systems with short-range interactions and ergodic behavior, gravitational systems are inherently non-extensive due to the presence of long-range interactions. This motivates the consideration of generalized entropy formalisms \cite{ent_tsallis, tsallisbook}.

In Tsallis statistics, the entropy is given by
\begin{equation}
    S_q = k \frac{1 - \sum_i p_i^q}{q - 1},
    \label{eq:tsallisent}
\end{equation}
where $q$ is the non-extensivity parameter, and the standard BG entropy is recovered in the limit $q \to 1$.

In the context of gravitational systems, it has been proposed that the entropy-area relation may be modified according to \cite{tsallis, saridakis3, tavayef}
\begin{equation}
    S = \gamma A^{\delta},
    \label{eq:tsallis}
\end{equation}
where $A \sim L^2$ is the horizon area, $\gamma$ is a constant, and $\delta$ parametrizes deviations from the standard Bekenstein--Hawking entropy, which is recovered for $\delta = 1$. This leads to the scaling
\begin{equation}
    S \propto L^{2\delta}.
    \label{eq:tsallis_l}
\end{equation}

Following the holographic principle, the ultraviolet (UV) and infrared (IR) cutoffs are related through the constraint that the total energy in a region of size $L$ should not exceed the mass of a black hole of the same size \cite{cohen}. This results in a vacuum energy density of the form
\begin{equation}
    \rho_{\Lambda} = B L^{2\delta - 4},
\end{equation}
where $B$ is a constant.

The parameter $\delta$ thus controls the deviation from the standard area-law scaling. While $\delta = 1$ reproduces the usual holographic dark energy scenario, values $\delta \neq 1$ correspond to modified entropy scaling. These deviations are often interpreted as effective non-extensive corrections associated with gravitational systems, although their precise microscopic origin remains unclear. In the present work, $\delta$ is treated as a phenomenological parameter, allowing us to isolate the impact of the infrared cutoff on cosmological observables.

\subsection{Background Cosmology}

We consider a spatially flat Friedmann--Lemaître--Robertson--Walker (FLRW) Universe, with metric
\begin{equation}
    ds^2 = -dt^2 + a^2(t)\, d\mathbf{x}^2,
    \label{eq:flrw}
\end{equation}
where $a(t)$ is the scale factor.

The background evolution is governed by the Friedmann equations
\begin{equation}
    3 \mpl^2 H^2 = \rho_{\mathrm{m}} + \rho_{\mathrm{DE}}
    \label{eq:fried1}
\end{equation}
\begin{equation}
    - 2 \mpl^2 \dot H = \rho_{\mathrm{m}} + p_{\mathrm{m}} + \rho_{\mathrm{DE}} + p_{\mathrm{DE}}
\end{equation}
where $\rho_{\mathrm{m}}$ and $p_{\mathrm{m}}$ denote the energy density and pressure of matter, while $\rho_{\mathrm{DE}}$ and $p_{\mathrm{DE}}$ correspond to dark energy.

We assume no interaction between dark matter and dark energy, so each component satisfies its own continuity equation,
\begin{equation}
    \dot \rho_{\mathrm{m}} + 3 H (\rho_{\mathrm{m}} + p_{\mathrm{m}}) = 0,
    \label{eq:contmat}
\end{equation}
\begin{equation}
    \dot \rho_{\mathrm{DE}} + 3 H (\rho_{\mathrm{DE}} + p_{\mathrm{DE}}) = 0.
    \label{eq:contde}
\end{equation}
The density parameters for matter and dark energy are defined as 
\begin{equation}
    \Omega_{\mathrm{m}} = \frac{\rho_{\mathrm{m}}}{3 \mpl^2 H^2}, \Omega_{\mathrm{DE}} = \frac{\rho_{\mathrm{DE}}}{3 \mpl^2 H^2}.
    \label{eq:dens}
\end{equation}
%For a spatially flat Universe with pressureless matter and HDE as dark energy, we can define the normalized Hubble parameter as \cite{salzano, naik}
For a spatially flat Universe, we have $\Omega_{\mathrm{m}} + \Omega_{\mathrm{DE}} = 1$. Assuming pressureless matter ($p_{\mathrm{m}} = 0$), the matter density evolves as $\rho_{\mathrm{m}} \propto a^{-3}$.

It is convenient to define the normalized Hubble parameter $E(a)$ such that \cite{salzano, naik}
\begin{equation}
    E^2(a) \equiv \frac{H^2(a)}{H^2_0} = \frac{\Omega_{\mathrm{m0}} a^{-3}}{1 - \Omega_{\mathrm{DE}}}
    \label{eq:Ea}
\end{equation}
where $\Omega_{\mathrm{m0}}$ is the present-day matter density parameter.

\begin{figure*}[]
    \centering
    \begin{subfigure}{0.45\textwidth}
        \centering
        \includegraphics[width=\linewidth]{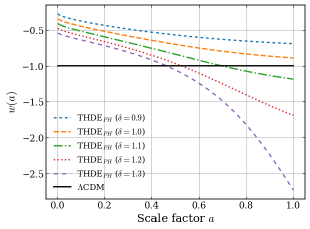}
        \caption{}
        \label{fig:wa_ph}
    \end{subfigure}
    \hspace{0.05\textwidth}
    \begin{subfigure}{0.45\textwidth}
        \centering
        \includegraphics[width=\linewidth]{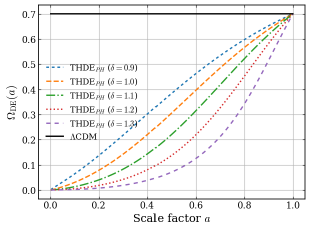}
        \caption{}
        \label{fig:omde_ph}
    \end{subfigure}
    
\vspace{0.5cm}

    \begin{subfigure}{0.45\textwidth}
        \centering
        \includegraphics[width=\linewidth]{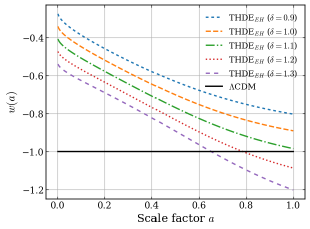}
        \caption{}
        \label{fig:wa_eh}
    \end{subfigure}
    \hspace{0.05\textwidth}
    \begin{subfigure}{0.45\textwidth}
        \centering
        \includegraphics[width=\linewidth]{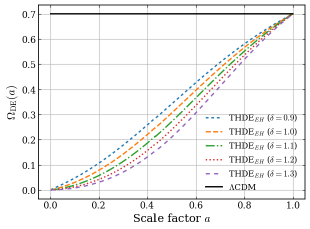}
        \caption{}
        \label{fig:omde_eh}
    \end{subfigure}
    
    \caption{The top left panel (Fig.~\ref{fig:wa_ph}) shows the evolution of $w(a)$ for THDE with the particle horizon as IR cutoff with different values of $\delta$ along with the $\Lambda$CDM model and the top right panel (Fig.~\ref{fig:omde_ph}) shows the evolution of $\Omega_{\mathrm{DE}}(a)$ for the same models. The bottom two panels (Fig.~\ref{fig:wa_eh} and Fig.~\ref{fig:omde_eh}) show the evolution of the same quantities for THDE with the future event horizon as IR cutoff, respectively. }
    \label{fig:one}
\end{figure*}

\begin{figure*}[]
    \centering
    \begin{subfigure}{0.45\textwidth}
        \centering
        \includegraphics[width=\linewidth]{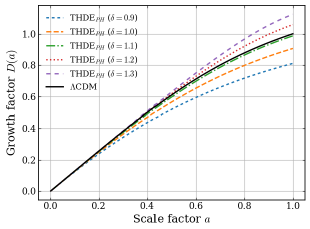}
        \caption{}
        \label{fig:da_ph}
    \end{subfigure}
    \hspace{0.05\textwidth}
    \begin{subfigure}{0.45\textwidth}
        \centering
        \includegraphics[width=\linewidth]{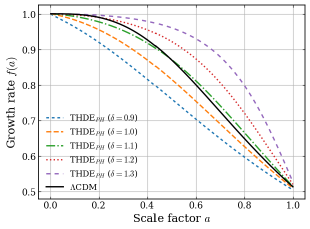}
        \caption{}
        \label{fig:fa_ph}
    \end{subfigure}

\vspace{0.5cm}

    \begin{subfigure}{0.45\textwidth}
        \centering
        \includegraphics[width=\linewidth]{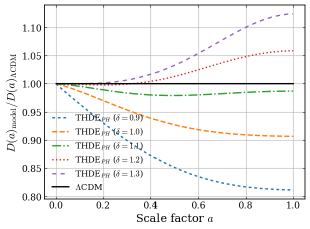}
        \caption{}
        \label{fig:rel_da_ph}
    \end{subfigure}
    \hspace{0.05\textwidth}
    \begin{subfigure}{0.45\textwidth}
        \centering
        \includegraphics[width=\linewidth]{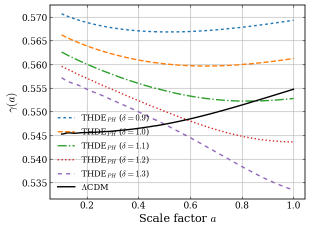}
        \caption{}
        \label{fig:gamma_ph}
    \end{subfigure}
    
    \caption{The top left and the top right panels (Fig.~\ref{fig:da_ph} and Fig.~\ref{fig:fa_ph}) show the evolution of $D(a)$ and $f(a)$ for THDE with the particle horizon as IR cutoff with different values of $\delta$ along with the $\Lambda$CDM model. Here, we set $D(a = 1) = 1$ for the $\Lambda$CDM model and plot $D(a)$ for other models with respect to it. The two bottom panels show the evolution of $D(a)_{\text{model}} / D(a)_{\Lambda\mathrm{CDM}}$ with the scale factor $a$ (Fig.~\ref{fig:rel_da_ph}) and the evolution of growth index $\gamma(a)$ with $a$ (Fig.~\ref{fig:gamma_ph}), respectively. } 
    \label{fig:two}
\end{figure*}

\subsection{Tsallis Holographic Dark Energy (THDE) Models}

We now specify the THDE model by choosing different infrared (IR) cutoffs, which determine the dynamical evolution of the dark energy density.

\subsubsection{Particle horizon}

We first consider the particle horizon as the IR cutoff. In the standard holographic dark energy framework, choosing the particle horizon does not lead to late-time accelerated expansion \cite{li}. However, it has been shown that within the Tsallis holographic framework, this choice can give rise to an accelerating Universe \cite{note}. It is therefore of interest to examine whether such a model remains viable when confronted with structure formation data.

The particle horizon $R_p$ is defined as 
\begin{equation}
    R_p = a(t) \int_0^t \frac{dt^{\prime}}{a(t^{\prime})} = a \int_0^a \frac{da^{\prime}}{H(a^{\prime}) a^{\prime 2}}
    \label{eq:ph}
\end{equation}
leading to $\rho_{\mathrm{DE}} = B R_p^{2 \delta - 4}$. The Hubble parameter can be constructed using Eq.~(\ref{eq:fried1}). Defining 
\begin{equation}
    F = \left(\frac{3 \Omega_{\mathrm{DE}} H^{2 \delta - 2}}{B}\right)^{1/(4 - 2 \delta)}
    \label{eq:F}
\end{equation}
and following \cite{note}, we obtain the differential equation of $\Omega_{\mathrm{DE}}$ and the equation of state $w(a)$ as 
\begin{equation}
    \frac{\Omega^{\prime}_{\mathrm{DE}}}{\Omega_{\mathrm{DE}} ( \Omega_{\mathrm{DE}} - 1)} = 1 + 2 (\delta - 2) F - 2 \delta
    \label{eq:ode_ph}
\end{equation}
and 
\begin{equation}
    w_{\mathrm{DE}} = - 1 - \left(\frac{2 \delta - 4}{3}\right) (1 + F)
    \label{eq:eos_ph}
\end{equation}
We plot the evolution of $\Omega_{\mathrm{DE}}(a)$ and $w(a)$ for some $\delta$ values in Fig.~\ref{fig:one} (top panels).

\subsubsection{Future event horizon}

Choosing the future event horizon as the IR cutoff to construct a THDE model is quite popular, since the future event horizon leads to a phase of late-time accelerating expansion of the Universe for the standard HDE \cite{li, wang2}. However, this leads to a few conceptual issues \cite{gaoricci, cruz, naik}. We consider HDE to be a phenomenological description of a component responsible for accelerating expansion in this work.  

The future event horizon $R_e$ is defined as 
\begin{equation}
    R_e = a(t) \int_t^{\infty} \frac{dt^{\prime}}{a(t^{\prime})} = a \int_a^{\infty} \frac{da^{\prime}}{H(a^{\prime}) a^{\prime 2}}
    \label{eq:eh}
\end{equation}
Accordingly, the dark energy density takes the form $\rho_{\mathrm{DE}} = B R_e^{2 \delta - 4}$. To construct the holographic dark energy model with the future event horizon as IR cutoff, we follow \cite{saridakis3}. The evolution equation of $\Omega_{\mathrm{DE}}$ is 
\begin{align}
    \frac{\Omega^{\prime}_{\mathrm{DE}}}{\Omega_{\mathrm{DE}} (1 - \Omega_{\mathrm{DE}})} &= 2 \delta - 1 + Q (1 - \Omega_{\mathrm{DE}})^{\frac{1 - \delta}{2 (2 - \delta)}} \Omega_{\mathrm{DE}}^{\frac{1}{2 (2 - \delta)}} e^{\frac{3 (1 - \delta)}{2 (2 - \delta)} x},
    \label{eq:ode_eh}
\end{align}
where 
\begin{equation}
    Q = 2 (2 - \delta) \left(\frac{B}{3 \mpl^2}\right)^{\frac{1}{2 (\delta - 2)}} \left(H_0 \sqrt{\Omega_{m0}}\right)^{\frac{1 - \delta}{\delta - 2}}
\end{equation}
and a prime over a quantity implies derivative with respect to $x = \ln a$. The equation of state is given by 
\begin{equation}
    w_{\mathrm{DE}} = \frac{1 - 2 \delta}{3} - \frac{Q}{3} \Omega_{\mathrm{DE}}^{\frac{1}{2 (2 - \delta)}} (1 - \Omega_{\mathrm{DE}})^{\frac{\delta - 1}{2 (\delta - 2)}} e^{\frac{3 (1 - \delta)}{2 (\delta - 2)} x}.
    \label{eq:eos_eh}
\end{equation}
$\Omega_{\mathrm{DE}}(a)$ and $w(a)$ evolutions for some $\delta$ values are plotted in Fig.~\ref{fig:one} (bottom panels).

% ==================================================
\section{Linear Growth of Matter Perturbations}
\label{sec:growth}

The evolution of large-scale structure in the Universe is governed by the growth of matter density perturbations. In the linear regime, where density fluctuations are small ($\delta \ll 1$), the growth of matter perturbations provides a powerful probe to distinguish between different dark energy models. We assume that dark energy perturbations are negligible on sub-horizon scales, as is commonly done in minimally coupled dark energy models with sound speed of order unity. Under this assumption, the growth of structure is governed solely by matter perturbations within general relativity. This assumption may be non-trivial in holographic dark energy models, where the energy density depends on non-local quantities such as horizon scales. A fully consistent treatment would require incorporating perturbations of the IR cutoff itself. In the present work, we adopt the standard smooth dark energy approximation, and our results should be interpreted within this framework.

We define the matter density contrast as
\begin{equation}
    \delta \equiv \frac{\delta \rho_m}{\rho_m}.
    \label{eq:delta}
\end{equation}
Assuming a pressureless matter component and neglecting anisotropic stress, the evolution of $\delta$ on sub-horizon scales is governed by the linear perturbation equation
\begin{equation}
    \ddot{\delta} + 2 H \dot{\delta} - 4 \pi G \rho_m \delta = 0.
    \label{eq:delta_t}
\end{equation}

It is convenient to rewrite this equation in terms of the scale factor $a$ as the independent variable. Using the relation $\frac{d}{dt} = H a \frac{d}{da}$, Eq.~(\ref{eq:delta_t}) can be rewritten as
\begin{equation}
    \delta''(a) + \left( \frac{H'(a)}{H(a)} + \frac{3}{a} \right)\delta'(a) - \frac{3}{2} \frac{\Omega_m(a)}{a^2} \delta(a) = 0,
    \label{eq:delta_a}
\end{equation}
where a prime denotes derivative with respect to $a$, and the matter density parameter evolves as
\begin{equation}
    \Omega_m(a) = \frac{\Omega_{m0} a^{-3}}{E^2(a)}.
    \label{eq:om_a}
\end{equation}
The growth of perturbations is thus determined by the background expansion through $H(a)$ and $\Omega_m(a)$, implying that different IR cutoffs affect structure formation indirectly via the modified expansion history.

To quantify the growth of structure, we define the growth factor $D(a)$ as
\begin{equation}
    D(a) = \frac{\delta(a)}{\delta(a=1)},
    \label{growth_fac}
\end{equation}
which is normalized to unity at the present epoch. The logarithmic growth rate is given by
\begin{equation}
    f(a) = \frac{d \ln D}{d \ln a}.
    \label{growth_rate}
\end{equation}
In a wide class of cosmological models where dark energy is smooth and the growth of structure occurs within General Relativity, the growth rate can be well approximated by the parametrization
\begin{equation}
    f(a) \approx \Omega_m(a)^{\gamma}
    \label{eq:gamma}
\end{equation}
where $\gamma$ is known as the growth index. For slowly evolving dark energy models, $\gamma$ is nearly constant but it can be time-depedent for more general scenarios. 

An important observable related to redshift-space distortions is the quantity $f\sigma_8(z)$, defined as
\begin{equation}
    f\sigma_8(z) = f(z)\, \sigma_8(z),
    \label{eq:fs8}
\end{equation}
where $\sigma_8(z)$ is the root-mean-square amplitude of matter fluctuations on scales of $8\,h^{-1}\,\mathrm{Mpc}$. Its redshift evolution is given by
\begin{equation}
    \sigma_8(z) = \sigma_{8,0} \frac{D(z)}{D(0)},
    \label{eq:s8_norm}
\end{equation}
where $\sigma_{8,0}$ is the present-day value.

For a given cosmological model, the evolution of $H(a)$ obtained from the background dynamics determines the behavior of $\Omega_m(a)$ and thus governs the growth of perturbations through Eq.~(\ref{eq:delta_a}). In the case of THDE, different choices of the infrared cutoff modify the expansion history, which in turn affects the growth of structure. To isolate the impact of IR cutoff on structure formation, we fix $\delta$ to some representative values. We consider five different values of the Tsallis parameter, $\delta = 0.9, 1.0, 1.1, 1.2,$ and $1.3$. The case $\delta = 1$ corresponds to the standard holographic dark energy model, while $\delta \neq 1$ parametrizes deviations from it. Observational studies typically constrain $\delta$ to be close to unity when treated as a free parameter \cite{mahnaz, saridakis3}, and the chosen values allow us to explore moderate deviations while remaining within observationally relevant ranges. Rather than performing a full parameter estimation, this approach enables a controlled comparison that highlights the role of the IR cutoff independently of entropy-driven effects. Thus, within the present framework, the impact of different IR cutoffs in THDE models enters the growth of structure through the modified expansion history. 

In this work, we numerically solve Eq.~(\ref{eq:delta_a}) for each model using appropriate initial conditions deep in the matter-dominated era, where the growing mode behaves as $\delta(a) \propto a$. We also set $\Omega_{m0} = 0.3$, $\Omega_{\mathrm{DE}0} = 0.7$, and $\sigma_{8, 0} = 0.811$ for all models. These solutions are then used to compute the growth factor $D(a)$, the growth rate $f(a)$, and the observable $f\sigma_8(z)$, which we compare with observational data in Sec.~\ref{sec:results}.

\subsection{Data}
\label{sec:data}

We use measurements of the quantity $f\sigma_8(z)$, which provide a direct probe of the growth of cosmic structure through redshift-space distortions. These data are taken from the compilation of \cite{leandros}, which includes measurements from multiple galaxy surveys spanning a wide range of redshifts.

To avoid potential biases due to correlated measurements, we construct a reduced dataset consisting only of statistically independent data points. This subset is listed in Table.~\ref{tab:rsd-clean} and is used in our statistical analysis. We assume Gaussian errors and neglect correlations among the selected data points.

To quantify the agreement between theoretical predictions and observational data, we construct the standard $\chi^2$ statistic,
\begin{equation}
    \chi^2 = \sum_i \frac{\left[f\sigma_8^{\mathrm{th}}(z_i) - f\sigma_8^{\mathrm{obs}}(z_i)\right]^2}{\sigma_i^2},
    \label{eq:chi_sq}
\end{equation}
where $f\sigma_8^{\mathrm{th}}(z_i)$ and $f\sigma_8^{\mathrm{obs}}(z_i)$ denote the theoretical prediction and observed value at redshift $z_i$, respectively, and $\sigma_i$ is the corresponding uncertainty.

To quantitatively compare different models, we use the Akaike Information Criterion (AIC) and the Bayesian Information Criterion (BIC), defined as
\begin{equation}
    \mathrm{AIC} = \chi^2_{\min} + 2k,
    \label{eq:aic}
\end{equation}
\begin{equation}
    \mathrm{BIC} = \chi^2_{\min} + k \ln N,
    \label{eq:bic}
\end{equation}
where $k$ is the number of free parameters and $N$ is the number of data points.

To assess the relative performance of a given model with respect to a reference model (taken here to be $\Lambda$CDM), we compute
\begin{equation}
    \Delta \mathrm{AIC} = \mathrm{AIC}_{\text{model}} - \mathrm{AIC}_{\Lambda\mathrm{CDM}},
\end{equation}
\begin{equation}
    \Delta \mathrm{BIC} = \mathrm{BIC}_{\text{model}} - \mathrm{BIC}_{\Lambda\mathrm{CDM}}.
\end{equation}

The relative differences $\Delta \chi^2$, $\Delta$AIC, and $\Delta$BIC provide a measure of model preference. Roughly speaking, $|\Delta \chi^2| \lesssim 1$ indicates statistically indistinguishable fits. For the information criteria, we adopt thresholds motivated by the Jeffreys scale (originally defined for Bayesian evidence and Bayes factors, and often applied to $\Delta \mathrm{BIC}$ as an approximation; see, for example, \cite{trotta}). According to this scale, $\Delta \mathrm{BIC} \lesssim 2$ suggest substantial support for a model, values in the range $4$--$7$ indicate less support, and values $\gtrsim 10$ imply strong disfavor compared to the reference model. We apply the same thresholds heuristically to $\Delta \mathrm{AIC}$ for ease of comparison. We explicitly compute $\Delta \mathrm{AIC}$ and $\Delta \mathrm{BIC}$ relative to the $\Lambda$CDM model to assess model preference.

\begin{table}[htbp]
\caption{Independent compilation of $f\sigma_8(z)$ measurements used in this work.}
\label{tab:rsd-clean}
\centering
\begin{tabular}{ccc}
\hline
Dataset & $z$ & $f\sigma_8(z)$ \\
\hline

6dFGS & 0.067 & $0.423 \pm 0.055$\\
SDSS-MGS & 0.15 & $0.490 \pm 0.145$ \\
GAMA & 0.18 & $0.360 \pm 0.090$ \\

SDSS-LRG & 0.35 & $0.440 \pm 0.050$ \\
SDSS-DR7-LRG & 0.35 & $0.429 \pm 0.089$ \\

WiggleZ & 0.44 & $0.413 \pm 0.080$ \\
WiggleZ & 0.60 & $0.390 \pm 0.063$ \\
WiggleZ & 0.73 & $0.437 \pm 0.072$ \\

BOSS DR12 & 0.38 & $0.497 \pm 0.045$ \\
BOSS DR12 & 0.51 & $0.458 \pm 0.038$ \\
BOSS DR12 & 0.61 & $0.436 \pm 0.034$ \\

VIPERS & 0.60 & $0.550 \pm 0.120$ \\
VIPERS & 0.86 & $0.400 \pm 0.110$ \\

VIPERS v7 & 0.76 & $0.440 \pm 0.040$ \\

FastSound & 1.40 & $0.482 \pm 0.116$ \\

SDSS-IV & 0.978 & $0.379 \pm 0.176$ \\
SDSS-IV & 1.23 & $0.385 \pm 0.099$  \\
SDSS-IV & 1.526 & $0.342 \pm 0.070$ \\
SDSS-IV & 1.944 & $0.364 \pm 0.106$ \\

2MTF & 0.001 & $0.505 \pm 0.085$ \\

6dFGS+SnIa & 0.02 & $0.428 \pm 0.0465$ \\

\hline
\end{tabular}
\end{table}

% ==================================================
\section{Results}
\label{sec:results}

\begin{figure*}[]
    \centering
    \begin{subfigure}{0.45\textwidth}
        \centering
        \includegraphics[width=\linewidth]{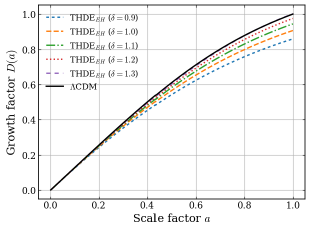}
        \caption{}
        \label{fig:da_eh}
    \end{subfigure}
    \hspace{0.05\textwidth}
    \begin{subfigure}{0.45\textwidth}
        \centering
        \includegraphics[width=\linewidth]{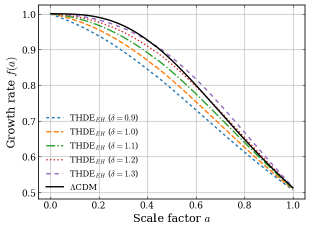}
        \caption{}
        \label{fig:fa_eh}
    \end{subfigure}

\vspace{0.5cm}

    \begin{subfigure}{0.45\textwidth}
        \centering
        \includegraphics[width=\linewidth]{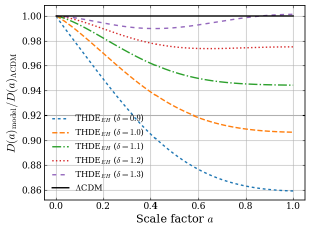}
        \caption{}
        \label{fig:rel_da_eh}
    \end{subfigure}
    \hspace{0.05\textwidth}
    \begin{subfigure}{0.45\textwidth}
        \centering
        \includegraphics[width=\linewidth]{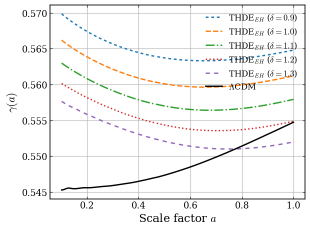}
        \caption{}
        \label{fig:gamma_eh}
    \end{subfigure}
    
    \caption{The top left and the top right panels (Fig.~\ref{fig:da_eh} and Fig.~\ref{fig:fa_eh}) show the evolution of $D(a)$ and $f(a)$ for THDE with the future event horizon as IR cutoff with different values of $\delta$ along with the $\Lambda$CDM model. Here, we set $D(a = 1) = 1$ for the $\Lambda$CDM model and plot $D(a)$ for other models with respect to it. The two bottom panels show the evolution of $D(a)_{\text{model}} / D(a)_{\Lambda\mathrm{CDM}}$ with the scale factor $a$ (Fig.~\ref{fig:rel_da_eh}) and the evolution of growth index $\gamma(a)$ with $a$ (Fig.~\ref{fig:gamma_eh}), respectively. }
    \label{fig:three}
\end{figure*}

\begin{figure*}
    \centering
    \begin{subfigure}{0.45\textwidth}
        \centering
        \includegraphics[width=\linewidth]{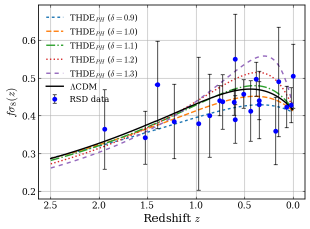}
        \caption{}
        \label{fig:fs8_ph}
    \end{subfigure}
    \hspace{0.05\textwidth}
    \begin{subfigure}{0.45\textwidth}
        \centering
        \includegraphics[width=\linewidth]{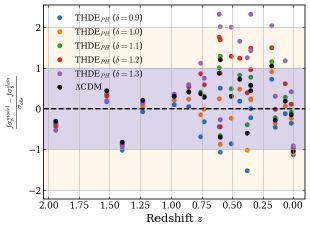}
        \caption{}
        \label{fig:res_ph}
    \end{subfigure}

    \vspace{0.5cm}

    \begin{subfigure}{0.45\textwidth}
        \centering
        \includegraphics[width=\linewidth]{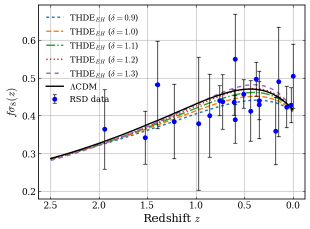}
        \caption{}
        \label{fig:fs8_eh}
    \end{subfigure}
    \hspace{0.05\textwidth}
    \begin{subfigure}{0.45\textwidth}
        \centering
        \includegraphics[width=\linewidth]{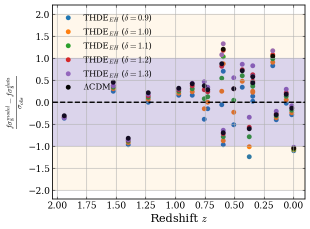}
        \caption{}
        \label{fig:res_eh}
    \end{subfigure}
    
    \caption{The top left panel (Fig.~\ref{fig:fs8_ph}) shows the plot of theoretical $f \sigma_8 (z)$ with $z$ for THDE with particle horizon as IR cutoff with different values of $\delta$ along with the same for the $\Lambda$CDM model and the observational data of $f\sigma_8(z)$ given in Table.~\ref{tab:rsd-clean}. The top right panels shows the normalised residuals $(f \sigma_8^{\text{theory}} - f \sigma_8^{\text{data}}) / \sigma_{obs}$ with $z$ for the same models. The two bottom panel show the same quantities for THDE with future event horizon as IR cutoff. }
    \label{fig:four}
\end{figure*}

\begin{table}[htbp]
\centering
\caption{Minimum $\chi^2$ values for THDE models with event horizon (EH) and particle horizon (PH) for different values of $\delta$, compared with $\Lambda$CDM. Since all models have the same number of free parameters, the corresponding $\Delta$AIC and $\Delta$BIC are identical to $\Delta \chi^2$. }
\label{tab:thde_chi2}
\begin{tabular}{lccccc}
\toprule
Model & $\delta$ & $\chi^2_{\min}$ & $\Delta \chi^2_{\min}$ & $\Delta$AIC & $\Delta$BIC\\

\midrule

$\Lambda$CDM & -- & 8.00 & 0.00 & 0.00 & 0.00\\

\midrule

THDE (EH) & 0.9 & 6.87 & -1.13 & -1.13 & -1.13\\
THDE (EH) & 1.0 & 6.38 & -1.62 & -1.62 & -1.62\\
THDE (EH) & 1.1 & 6.76 & -1.24 & -1.24 & -1.24\\
THDE (EH) & 1.2 & 7.90 & -0.10 & -0.10 & -0.10\\
THDE (EH) & 1.3 & 9.71 & 1.71 & 1.71 & 1.71\\

\midrule

THDE (PH) & 0.9 & 8.69 & 0.69 & 0.69 & 0.69\\
THDE (PH) & 1.0 & 6.38 & -1.62 & -1.62 & -1.62\\
THDE (PH) & 1.1 & 9.16 & 1.16 & 1.16 & 1.16\\
THDE (PH) & 1.2 & 18.86 & 10.86 & 10.86 & 10.86\\
THDE (PH) & 1.3 & 35.42 & 27.42 & 27.42 & 27.40\\

\bottomrule
\end{tabular}
\end{table}

We present the results of our analysis in Figs.~\ref{fig:one}--\ref{fig:four}. To isolate the role of the infrared (IR) cutoff, we first discuss the behaviour of the THDE models constructed using the particle horizon (PH) and the future event horizon (EH) separately. We then compare their predictions with the $\Lambda$CDM model and observational data. Since we do not perform a full parameter estimation and instead consider fixed representative values of $\delta$, the statistical comparison presented here should be regarded as indicative rather than a complete likelihood analysis.

\subsection{Particle Horizon as IR Cutoff}

Fig.~\ref{fig:one} (top panels) shows the evolution of the equation of state $w(a)$ and the dark energy density parameter $\Omega_{\mathrm{DE}}(a)$ for THDE models with the particle horizon as the IR cutoff. A key feature that emerges is the strong dependence of the background evolution on the Tsallis parameter $\delta$.

While all models converge to the same asymptotic limits at early and late times, the timing of dark energy domination differs significantly. In particular, for larger values of $\delta$ (e.g., $\delta = 1.2$ and $1.3$), $\Omega_{\mathrm{DE}}$ becomes dynamically important much later compared to smaller values such as $\delta = 0.9$ and $1.0$. This delay directly impacts the growth of structure.

A similar trend is observed in the evolution of the equation of state. For $\delta = 1.2$ and $1.3$, $w(a)$ exhibits a strong phantom behaviour and deviates significantly from the $\delta = 0.9$, $1.0$ and $1.1$ cases. This indicates that the effective dark energy component is not only dynamically delayed but also exhibits a qualitatively different evolution. Overall, the particle horizon models show large variations in background quantities with changing $\delta$, reflecting the sensitivity of these models to the entropy scaling.

The consequences of this delayed dark energy domination are clearly visible in the growth of structure. In Fig.~\ref{fig:two}, we show the evolution of the growth factor $D(a)$ and the growth rate $f(a)$. For $\delta = 1.2$ and $1.3$, both $D(a)$ and $f(a)$ are enhanced relative to the $\Lambda$CDM model, while for $\delta = 0.9$ and $1.0$, the growth is suppressed. The case $\delta = 1.1$ remains closest to $\Lambda$CDM.

This behaviour can be directly traced to the evolution of $\Omega_{\mathrm{DE}}(a)$. For larger $\delta$, the delayed onset of dark energy domination allows matter perturbations to grow more efficiently over an extended period, leading to enhanced clustering. Conversely, for smaller $\delta$, earlier dark energy influence suppresses growth.

To highlight deviations from $\Lambda$CDM more clearly, we plot the relative growth factor $D(a)_{\text{model}}/D(a)_{\Lambda\mathrm{CDM}}$ in Fig.~\ref{fig:rel_da_ph}. We find that for $\delta = 0.9$, $1.0$, $1.2$, and $1.3$, the deviation from unity appears early and increases with time, whereas for $\delta = 1.1$, the ratio remains close to unity throughout the evolution. This confirms that only a narrow range of $\delta$ values can reproduce $\Lambda$CDM-like growth.

A particularly important diagnostic is the growth index $\gamma(a)$, shown in Fig.~\ref{fig:gamma_ph}. Unlike the $\Lambda$CDM model, where $\gamma(a)$ increases monotonically, the THDE model with the particle horizon shows two distinct behaviours depending on the value of $\delta$. For $\delta = 0.9$, $1.0$ and $1.1$,  $\gamma(a)$ initially decreases, indicating a phase of enhanced growth, followed by a late-time increase as dark energy begins to suppress structure formation. However, for larger values $\delta = 1.2$ and $1.3$, $\gamma(a)$ decreases monotonically throughout the evolution, implying sustained enhancement of growth without a clear transition to the standard suppression regime. For higher values of $\delta$, the onset of dark energy domination and the associated suppression of structure growth moves toward a larger value of scale factor, confirming the trend observed in the plot of $\Omega_{\mathrm{DE}}(a)$. For $\delta = 1.3$, this epoch occurs in the future, explaining the monotonic nature of $\gamma(a)$ in this case. 

Overall, we find that only specific choices of $\delta$ (notably $\delta \approx 1.1$) yield growth histories comparable to $\Lambda$CDM, while larger values lead to substantial deviations.

\subsection{Future Event Horizon as IR Cutoff}

We now turn to the case where the future event horizon is used as the IR cutoff. The corresponding evolution of $w(a)$ and $\Omega_{\mathrm{DE}}(a)$ is shown in Fig.~\ref{fig:one} (bottom panels).

In contrast to the particle horizon case, the background evolution here shows only mild dependence on $\delta$. The evolution of $\Omega_{\mathrm{DE}}(a)$ is nearly identical for all values of $\delta$, and dark energy becomes dynamically relevant at approximately the same epoch in all cases. Similarly, the equation of state remains close to $w \approx -1$ at late times, with only small deviations.

This uniformity in the background evolution leads to a correspondingly stable growth history. As shown in Fig.~\ref{fig:three}, the growth factor $D(a)$ and growth rate $f(a)$ for all values of $\delta$ remain close to those of the $\Lambda$CDM model. Despite the small deviations, a systematic trend can be observed in this case. The growth of structures is the smallest for the smallest value of $\delta$, $\delta = 1.3$ being the closest to the $\Lambda$CDM curves. 

The relative growth factor $D(a)_{\mathrm{model}} / D(a)_{\Lambda\mathrm{CDM}}$ shows the same systematic dependence on $\delta$. The case $\delta = 1.3$ remains closest to unity, while smaller values of $\delta$ exhibit increasing deviations from $\Lambda$CDM, indicating progressively stronger departures in the growth history. In all cases, $\gamma(a)$ shows an initial decrease followed by a gradual increase at late times, indicating a transient phase of enhanced growth before the onset of suppression. This behaviour is consistent with the relatively mild deviations observed in the growth factor and reflects the more stable evolution of dark energy in these models. 

This behaviour can be understood physically: since dark energy domination occurs at a similar epoch for all values of $\delta$, the duration of efficient matter clustering is nearly unchanged. As a result, the growth of structure closely tracks that of $\Lambda$CDM.

\subsection{Comparison with Observations}

In Fig.~\ref{fig:four}, we compare the theoretical predictions for $f\sigma_8(z)$ with observational data. We also show the normalised residuals, defined as $\frac{f\sigma_8^{\text{model}} - f\sigma_8^{\text{data}}}{\sigma_{\text{obs}}}$ to highlight systematic deviations.

For THDE models with the future event horizon as the IR cutoff, the theoretical predictions lie within the observational uncertainty bands across most of the redshift range. Only small deviations are observed, indicating that these models provide a fit comparable to the $\Lambda$CDM model.

In contrast, particle horizon models show significant deviations from the data, particularly for larger values of $\delta$. The $\delta = 1.3$ model exhibits strong overprediction of $f\sigma_8$ at low redshift, leading to large residuals and poor agreement with observations. This is consistent with the enhanced growth seen in the evolution of $D(a)$ and $f(a)$.

These results are quantitatively reflected in the $\chi^2$ values summarized in Table.~\ref{tab:thde_chi2}. We find that THDE models with the event horizon cutoff and $\delta \approx 0.9$--$1.3$ yield $\Delta$AIC values $\le 2$, indicating statistical consistency with $\Lambda$CDM within the present fixed-parameter comparison. On the other hand, particle horizon models with $\delta \geq 1.2$ are strongly disfavored, with large $\Delta\chi^2$ values.

%==============================================
\section{Discussion}
\label{sec:discussion}

We investigate the impact of the infrared (IR) cutoff on the growth of cosmic structures within THDE models. Our analysis demonstrates that, although different choices of IR cutoff can lead to qualitatively similar background evolution, their imprints on structure formation are markedly different.

A consistent physical picture emerges when the results are interpreted in terms of the timing of dark energy domination. The growth of cosmic structures is governed by the competition between gravitational clustering and the accelerated expansion driven by dark energy. The competition is controlled jointly by the IR cutoff and the Tsallis parameter $\delta$, which together determine how and when dark energy becomes dynamically significant. Different choices of the IR cutoff modify the evolution of the dark energy density and hence the Hubble parameter $H(a)$. This, in turn, changes the timing and duration of dark energy domination. If dark energy becomes dynamically important at a later epoch, matter perturbations have more time to grow, leading to enhanced structure formation. Conversely, an earlier onset of dark energy suppresses the growth of perturbations. Therefore, even in the absence of dark energy perturbations, different IR cutoffs can lead to distinct growth histories through their impact on the background expansion.

In models based on the particle horizon, the onset of dark energy domination is delayed, particularly for larger values of $\delta$. As a result, matter perturbations experience an extended phase of efficient gravitational clustering. This leads to enhanced growth, reflected in larger values of the growth factor $D(a)$, the growth rate $f(a)$, and the observable $f\sigma_8(z)$. However, this delayed suppression also results in a systematic overprediction of structure formation, causing significant tension with observational data for $\delta \gtrsim 1.2$.

In contrast, when the future event horizon is used as the IR cutoff, dark energy domination occurs at a more uniform and earlier epoch across different values of $\delta$. This results in a timely suppression of growth, preventing the prolonged clustering phase seen in particle horizon models. Consequently, the growth history remains close to that of the $\Lambda$CDM model, and these models provide a fit to the observational data comparable to $\Lambda$CDM.

An important result is the behaviour of the growth index $\gamma(a)$, which provides a deeper insight into the growth dynamics beyond what is captured by $D(a)$ and $f(a)$ alone. While the $\Lambda$CDM model exhibits a smooth and monotonic evolution of $\gamma(a)$, THDE models with particle horizon cutoff display a distinctly non-monotonic behaviour. In particular, the presence of a dip in $\gamma(a)$ signals a transition epoch during which dark energy begins to influence the dynamics but has not yet effectively suppressed the growth of perturbations.

This can be understood by noting that the growth index parametrizes the relation $f(a) = \Omega_m(a)^\gamma$. A decrease in $\gamma$ corresponds to an enhancement in the growth rate for a given matter density. Thus, the dip in $\gamma$ directly reflects the phase of enhanced clustering caused by the delayed onset of dark energy domination. At later times, as dark energy becomes dominant, $\gamma$ increases, indicating the onset of growth suppression. This transition is absent in $\Lambda$CDM, where the suppression of growth is more gradual and monotonic.

The non-monotonic evolution of $\gamma(a)$ therefore provides a clear diagnostic to distinguish THDE models from $\Lambda$CDM, even in cases where the background expansion history is nearly identical. This demonstrates that growth-based observables can effectively break degeneracies that are otherwise present at the level of $H(a)$.

From a statistical perspective, we find that THDE models with the future event horizon as the IR cutoff and $\delta \approx 1.0$--$1.2$ yield fits to the $f\sigma_8$ data that are statistically comparable to $\Lambda$CDM based on $\chi^2$ comparison with $f\sigma_8(z)$ data. In contrast, particle horizon models with $\delta \geq 1.2$ are strongly disfavored by current observations of $f\sigma_8(z)$. This indicates that structure formation provides a stringent constraint on the viability of these models. We highlight here that this result is obtained keeping $\delta$ fixed to some representative values, as our goal is to isolate the impact of the IR cutoff. Treating $\delta$ as a free parameter would obscure this comparison. A more rigorous analysis would treat $\delta$ as a free parameter and employ MCMC techniques using multiple independent datasets to estimate the value of $\delta$. In this way, we can also test whether these models can alleviate recent cosmological tensions such as the $H_0$ tension or the $S_8$ tension. We leave that for future work. 

Recent analyses of DESI BAO data, particularly in combination with other cosmological probes, have suggested a mild preference for dynamical dark energy models exhibiting phantom crossing \cite{desi_dr1, desi_dr2}. In the present THDE framework, phantom behavior can also arise for certain choices of the Tsallis parameter $\delta$, especially in models based on the particle horizon cutoff, as shown in Fig.~\ref{fig:one}. However, the evolution of the equation of state in these models differs from the behavior often inferred in phenomenological ($w_0$, $w_a$) parametrizations fitted to DESI data. In particular, the THDE models considered here typically evolve from quintessence-like behavior at earlier times toward phantom behavior at late times, whereas DESI-inspired ($w_0$, $w_a$) reconstructions may favor the opposite trend. Therefore, while THDE models can exhibit phantom crossing, a dedicated analysis would be required to determine whether they can reproduce the detailed redshift dependence suggested by DESI observations. We leave such an investigation for future work.

We emphasize that the parameter $\delta$ plays a central physical role in this framework. It governs the scaling of the dark energy density with the IR cutoff and thereby controls both the timing of dark energy domination and the effective stiffness of the cosmic fluid. Physically, $\delta$ can be interpreted as encoding deviations from standard extensive entropy, with larger values corresponding to stronger long-range correlations in the gravitational system. These correlations delay the onset of dark energy dominance and enhance structure formation, particularly in particle horizon models.

Finally, we note an important conceptual limitation of the present analysis. In THDE models, the dark energy density depends on non-local quantities such as the particle or event horizon, which are defined through integrals over the cosmic history. However, in our perturbation analysis, dark energy is treated as a smooth component with negligible fluctuations on sub-horizon scales. This approximation is commonly adopted in holographic dark energy studies. A fully consistent treatment would require incorporating perturbations of the non-local IR cutoff itself, possibly within an effective field theory framework. Alternatively, one can employ other cutoffs that are local, such as Ricci cutoff \cite{gaoricci} and Granda--Oliveros cutoff \cite{go1, go2, oliveros}. It would also be interesting to see how the results of this work change when interactions between HDE and dark matter are allowed. 

Overall, the results indicate that the infrared cutoff not only determines the background evolution but also imprints a characteristic signature on the growth history of the Universe. This suggests that structure formation can serve as a probe of the underlying non-extensive entropy formalism itself.

% ==================================================
\section*{Acknowledgments}

We thank an anonymous referee for their careful reading of the manuscript and constructive comments that have helped to improve the presentation of our results. This work has been substantially improved by some really important comments on an earlier draft by Biswajit Pandey and Rahul Shah. We also thank Debarun Paul and Tanmoy Paul for useful discussions. Finally, we acknowledge financial support from CSIR, Govt. of India through research associateship program (File No: 09/0093(19416)/2024-EMR-I).  

\section*{Data Availability}

The data underlying this study are contained in Table~\ref{tab:rsd-clean}. The compilation from which the independent subset was constructed is publicly available in Ref.~\cite{leandros}.

\bibliographystyle{apsrev4-2}
\bibliography{reference}

\end{document}